**Response to Comment on (A novel two-dimensional boron-carbon-nitride BCN monolayer: A First-principles insight [JAP21-AR-03574R])**

A. Bafekry*[1], M. Naseri[2], M. M. Fadlallah[3], I. Abdolhosseini Sarsari[4], M. Faraji[5], A. Bagheri Khatibani[6], M. Ghergherehchi[7], and D. Gogova[8]

1 Department of Radiation Application, Shahid Beheshti University, Tehran 1983969411, Iran

2 Department of Physics, Kermanshah Branch, Islamic Azad University, 6718997551 Kermanshah, Iran

3 Department of Physics, Faculty of Science, Benha University, 13518 Benha, Egypt

4 Department of Physics, Isfahan University of Technology, Isfahan 84156-83111, Iran

5 Micro and Nanotechnology Graduate Program, TOBB University of Economics and Technology, Sogutozu Caddesi No 43 Sogutozu, 06560 Ankara, Turkey

6 Nano Research Lab, Lahijan Branch, Islamic Azad University, 1616 Lahijan, Iran

7 Department of Electrical and Computer Engineering, Sungkyunkwan University, 16419 Suwon, South Korea

8 Department of Physics, Chemistry and Biology, Linkoping University, 581 83 Linköping, Sweden

Mortazavi et al. comment on the results of our paper "A novel two-dimensional boron-carbon-nitride BCN monolayer: A First-principles insight," in which we have considered different configurations of BCN. For clarification, we applied the fingerprint theory [1-3] to examine the similarity between the distinct structures. The fingerprint function is a crystal structure descriptor, an 1D-function related to the pair correlation function and diffraction patterns. It does not depend on absolute atomic coordinates, but only on interatomic distances. Small deviations in atomic positions will influence the fingerprints only slightly [1-3]. Fingerprint theory allows quantification of the degree of order and complexity of a crystal structure.

E-mail: bafekry.asad@gmail.com

Moreover, one could also measure the similarity between structures. For each given structure, one may also compute its own quasi-entropy as a measure of disorder and complexity of that structure.

$$S_{\mathrm{str}} = -\Sigma_A \frac{N_A}{N} < \ln\left(1 - D_{A_i A_j}\right)$$

where distances DAiAj are measured between the fingerprints of all ith and jth sites occupied by chemical species A, and the total quasi-entropy is the weighted sum over all chemical species, which measure, the physical difference between positions. In Table 1 one can see the structures with same quasi entropy has same structure, but it does not affect the paper's results and conclusions because we have focused on the structure No. 9.

| System | Quasi entropy |
|---|---|
| 1 | 0.1054 |
| 9 | 0.1055 |
| 12 | 0.1036 |
| 2 | 0.0000 |
| 10 | 0.0000 |
| 11 | 0.2111 |
| 3 | 0.0000 |
| 7 | 0.0000 |
| 4 | 0.1036 |
| 5 | 0.1036 |

Table 1: The quasi-entropy of the different configuration of the BCN structure based on the fingerprints theory [1-3].

The atomic structure of the BCN monolayer with configuration 9 in top view is depicted in Figure. 1. In the BCN, the graphene-like honeycomb lattice with a hexagonal primitive unit cell is formed by eight atoms (2B, 4 C, and 2 N). The honeycomb lattice of BCN belongs to the P3m1 space group with a unit cell lattice constant of 4.34 Å.

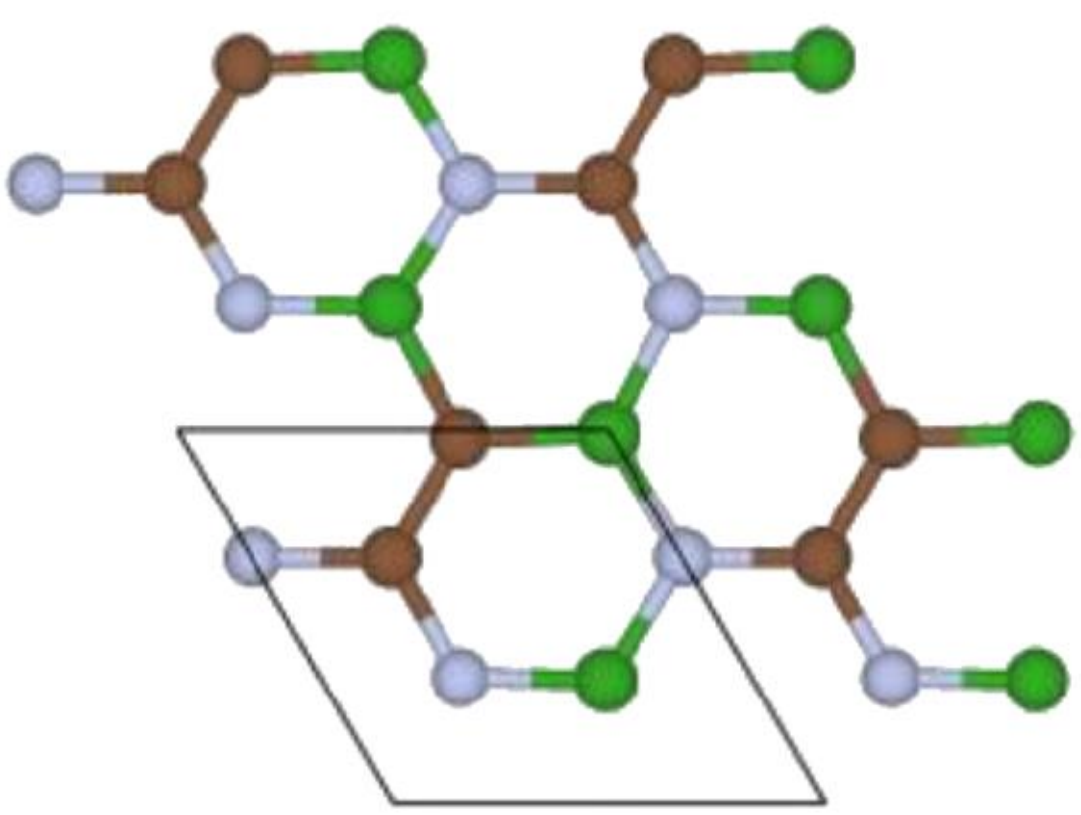

Figure 1: Configuration 9 of the BCN monolayer.

There are many approximations in the computational methods or models, including limitations for both experimental and computational studies. One should try to use the best suitable tool. For the predicted BCN structure we have calculated the bandgap using the PBE and HSE06 functionals. The PBE gives an underestimated band gap value, however, the HSE06 is a more accurate method. Using the HSE06 hybrid functional is common in the computational community because of its accuracy in predicting the band gap and band structures. Figure. 2 illustrates the reputation of these methods [4] in predicting the band gap and band structures of different materials.

For the application of the proposed structure (not only BCN structure) theory can also guide the experiments. In many cases it gives deeper insights and offers promising material systems candidates for the experimental peers to explore. For the case of BCN monolayer, although our theoretical calculations show that it exhibits a bandgap of 1.6 eV, which lies in the visible region of spectrum, we can claim that it may have promising applications such as electron emitters photovoltaic devices, and electro-catalysts .

To present more accurate information, it is more reasonable always to pay attention to the computational/experimental accuracy. However, nothing is perfect.

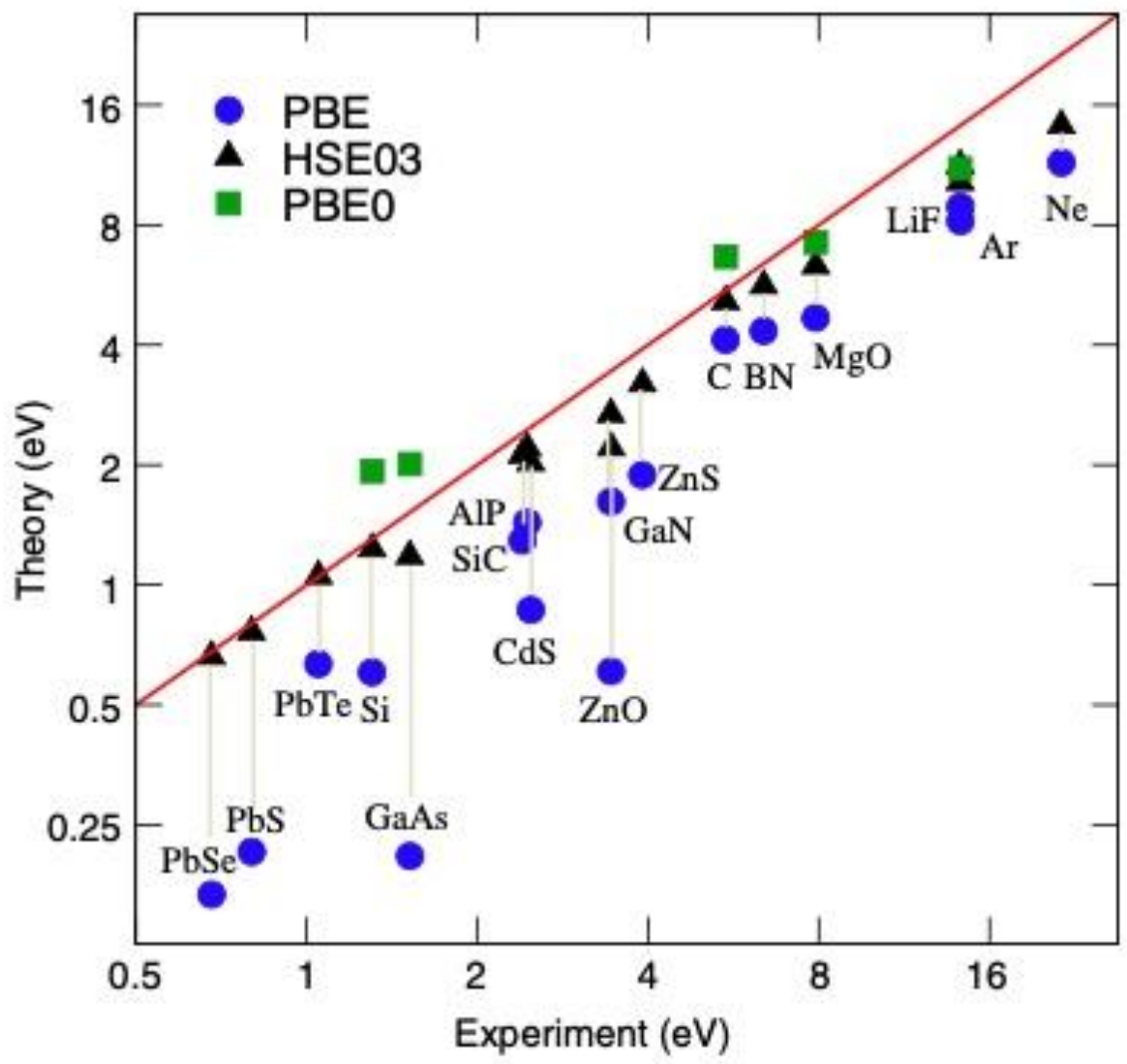

Figure 2: Band gaps from PBE, PBE0, and HSE03 calculations, plotted versus the experimental data

For the graphical presentation of the ELF calculation, we scaled the values between 0 and 0.5 to better visualize the ELF. Our ELF results are accurate, while those of Mortazavi are within 0-0.9

Furthermore, the difference charge density of BCN show accumulation and depletion on different atoms.

In the phonon calculations, the symmetrical points of the first Brillouin zone (BZ) are defined correctly in our paper. We chose the first BZ for the hexagonal BCN structure. Our phonon calculation shows no existence of any negative mode. It is better the frequency axis to be in arbitrary units when there are no experimental results yet. However, here we are aiming to show the absence of negative frequencies and it is properly done. We mentioned that unit of the Y-axis in the phonon band spectrum is $Cm^{-1}$.

There is decimal point put on a wrong place for the elastic parameter. The elastic constants are: C11=1389 GPa, C12=289 GPa, C13=18 GPa, C22=1403 GPa, C33 = 72 GPa, C 44=549 GPa, C 55= C66= 23 GPa. Thus, the Young’s modulus of BCN becomes 1328.86 GPa, which value is close to the experimental and corresponding theoretical values (1050-2400 GPa) [5-8]. For the elastic constants C13 and C33, which are uniaxial elastic constants along the z axis, we have calculated very small values, demonstrating that the interaction along the z axis is minimized, and therefore, the material studied behaves as a perfect 2D structure.

References:

[1] Oganov A.R., Valle M. (2009), How to quantify energy landscapes of solids, J. Chem. Phys. 130, 104504.

[2] Lyakhov A.O., Oganov A.R., Valle M. (2010), How to predict very large and complex crystal structures, Comp. Phys. Comm. 181, 1623-1632.

[3] Valle M., Oganov A.R. (2010), Crystal fingerprints space. A novel paradigm to study crystal structures sets, Acta Cryst. A66, 507-517.

[4] M. Marsman et al., J. Phys.: Condens. Matter 20, 064201 (2008)

[5] J.-U. Lee, D. Yoon, and H. Cheong, Estimation of Young's Modulus of Graphene by Raman Specroscopy. Nano Lett. 2012, 12, 9, 4444–4448.

[6] J. W. Jiang, J. S. Wang and B. W. Li, Phys. Rev. B: Condens. Matter Mater. Phys. 80, 113405 (2009).

[7] E. Konstantinova, S. O. Dantas, and P. M. V. B. Barone, Phys. Rev. B 74, 035417 (2006).

[8] F. Liu, P. Ming, and Ju Li, Phys. Rev. B 76, 064120 (2007).